\documentclass[journal=jctcce,manuscript=article,hyphens,maxauthors=60]{achemso}

\usepackage[version=3]{mhchem} 
\usepackage{xcolor}

\definecolor{oxygen}{RGB}{255,0,0}    
\definecolor{sulfur}{RGB}{255,200,0}  
\definecolor{nitrogen}{RGB}{0,0,255}  

\usepackage{colortbl} 
\usepackage{graphicx} 

\usepackage[T1]{fontenc} 
\usepackage[utf8x]{inputenc}
\usepackage{tabularx}
\usepackage{adjustbox}
\usepackage{lipsum}
\usepackage{longtable}
\usepackage{booktabs}
\usepackage{hyperref}
\usepackage{booktabs}
\usepackage{siunitx}
\usepackage[table]{xcolor}
\definecolor{lightcyan}{rgb}{0.88,1,1}
\usepackage{multirow}
\usepackage{braket}
\usepackage{amsfonts}
\usepackage{float}
\usepackage{threeparttable}
\usepackage{graphicx}
\usepackage{array}
\usepackage{mathtools}
\newcommand{\Lower}[1]{\smash{\lower 1.5ex \hbox{#1}}}

\usepackage{soul}
\usepackage[normalem]{ulem}

\author{Ram Dhari Pandey}
\affiliation{Institute of Physics, Faculty of Physics, Astronomy, and Informatics, Nicolaus Copernicus University in Toru\'{n}, Grudzi\k{a}dzka 5, 87-100 Toru\'{n}, Poland}
\author{Marta Ga\l{}y\'{n}ska}
\affiliation{Faculty of Chemistry, Nicolaus Copernicus University in Toru\'{n}, Grudzi\k{a}dzka 5, 87-100 Toru\'{n}, Poland}

\email{marta.galynska@umk.pl}
\author{Katharina Boguslawski}
\affiliation[Unknown University]
{Institute of Physics, Faculty of Physics, Astronomy, and Informatics, Nicolaus Copernicus University in Toru\'{n}, Grudzi\k{a}dzka 5, 87-100 Toru\'{n}, Poland}

\author{Pawe\l{} Tecmer}
\affiliation[Unknown University]
{Institute of Physics, Faculty of Physics, Astronomy, and Informatics, Nicolaus Copernicus University in Toru\'{n}, Grudzi\k{a}dzka 5, 87-100 Toru\'{n}, Poland}
\email{ptecmer@fizyka.umk.pl}

\title
  {Tuning Domain-Based Charge Transfer in Organic Dyes: Impact of Heteroatom Doping in the $\pi$-linker of Carbazole-Based Systems}

\begin{document}
\begin{tocentry}
\centering
\includegraphics[]{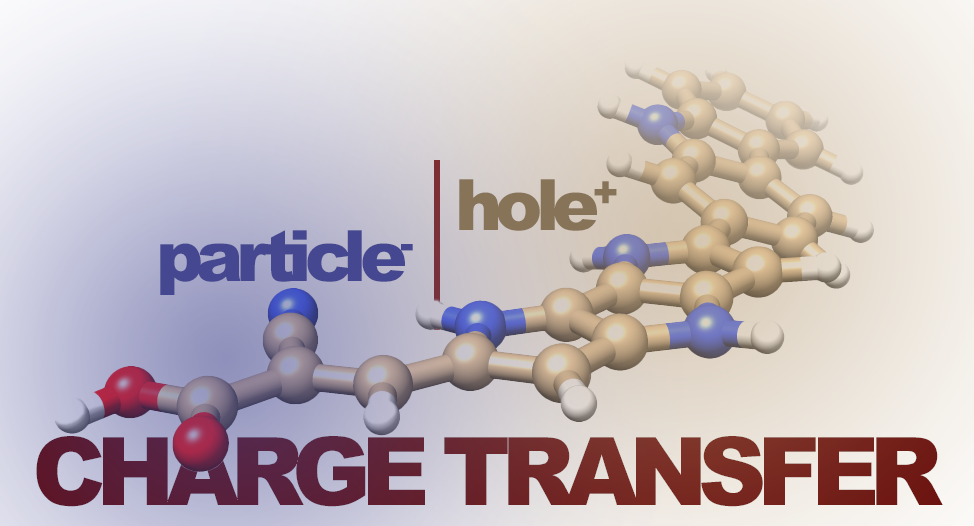}
\end{tocentry}
\begin{abstract}
This work presents an innovative computational study of domain-based charge transfer that leverages the localized orbitals of pair Coupled Cluster Doubles (pCCD).
This method enables both directional monitoring and quantitative assessment of charge transfer among donor (D), bridge (B), and acceptor (A) moieties.
We applied this approach to a series of newly designed carbazole-based prototypical organic dyes, doping the bridge at positions 1, 2, and 3 with nitrogen, oxygen, and sulfur atoms to generate mono-, di-, and tri-doped variants.
Our results demonstrate a clear and progressive enhancement in charge transfer as the degree of nitrogen or oxygen doping increases from mono- to di- to tri-doped systems.
For mono-doped dyes, the highest forward charge transfer from donor to bridge to acceptor (D$\xrightarrow{}$B$\xrightarrow{}$A) occurs when a heteroatom (N or O) is placed in the terminal ring of the bridge, closer to the acceptor.
In di-doped dyes, the largest forward charge transfer is observed when heteroatoms occupy both terminal positions, with one atom (N or S) adjacent to the donor and the other (N) near the acceptor.
Nitrogen-doped systems consistently outperform their oxygen and sulfur counterparts.
Among all variants, the organic dye doped with three nitrogen atoms at the bridge exhibits the most efficient and highest directional donor-to-acceptor charge transfer (42.6\%), making it the most promising candidate for potential applications in dye-sensitized solar cells. 
Finally, our calculations predict weak charge separation in all systems, indicating that charge transfer predominantly occurs from the bridge to the acceptor. 
\end{abstract}

\section{Introduction}
The demand for sustainable energy sources is increasing as a result of the growing population in the world, rising energy consumption, and environmental issues caused by the emission of harmful gases into the atmosphere by the combustion of fossil fuels.~\cite{emission1, emission2,emission3, emission4, emission5}
Solar energy is widely recognized as a substitute for a renewable and environment friendly energy source.~\cite{re-energy} Silicon-based solar cells offer stability and durability.
However, the complex purification process and high manufacturing cost limit its applications.~\cite{cost1,cost2,cost3}
Therefore, dye-sensitized solar cells (DSSCs) have emerged as a versatile and cost-effective alternative.~\cite{dsscs}
Sensitizers can be divided into two main categories. Ruthenium-based metal complexes are the most studied category of sensitizers, delivering photon-to-current conversion efficiency (PCE)~\cite{Ru-efficiency} around 12\%.
However, these complexes are associated with many drawbacks, such as complicated synthesis and purification processes in the laboratory,~\cite{Ru-synthesis} concerns about toxicity,~\cite{toxic1} and the high cost of ruthenium being a rare transition metal (low natural abundance) may restrict future large-scale applications.
In contrast, metal-free organic sensitizers emerge as the most promising candidate that offers a variety of advantages, including a facile synthesis process~\cite{synthesis,pi-linker2}, cost-effectiveness, flexibility to customize desired molecules for photophysical properties,~\cite{flexibility,photophysical-properties} better environmental sustainability and high absorption coefficients.~\cite{sustainability}
A PCE of up to 37\% has been reported for indoor photovoltaic applications.~\cite{efficiency}

A widely adopted and promising strategy for designing the most efficient metal-free organic sensitizers is based on the donor-$\pi$-conjugated-bridge-acceptor (D-B-A) model.
In this model, the choice for the most prominent and optimal organic electron donor (D) moieties, such as carbazole,~\cite{carbazole1, carbazole2, carbazole3} dimethylamino,~\cite{dma} triphenylamine,~\cite{tpp1, tpp2, tpp3}  coumarin,~\cite{coumarin1, coumarin2, coumarin3} indoline,~\cite{indoline1,indoline2,indoline3,indoline4} and phenothiazine~\cite{phenothiazine1,phenothiazine2,phenothiazine3}---is primarily governed by the capacity of an electron-donating group.
The most crucial and extensively studied $\pi$-linkers or bridges (B) include fused benzene systems, thiophene,~\cite{thiophene1,thiophene2} perylene,~\cite{perylene1,perylene2} benzothiadiazole,~\cite{benzothia1,benzothia2} quinoline,~\cite{quinoline1} and enediyne.~\cite{enediyne}
In the end, cyanoacrylic acid is the most commonly reported anchoring group in the literature.~\cite{anchoring-group,anchoring-group2,anchoring-group3} 
It emerges as a versatile acceptor (A).
This is because of the strong ability to withdraw electrons from the donor moiety of the dye. It especially withdraws electrons via the -CN group.
It also has the binding affinity of the carboxylic group (-COOH) with \ce{TiO2}.~\cite{binding-with-TiO2}
This binding allows the injection of photoexcited electrons~\cite{electron-ejection} from the lowest unoccupied molecular orbital (LUMO) of the dye.
The electrons are injected into the conduction band of the semiconductor, which must be present energetically at a lower level than that of the LUMO of the organic dye. The planarity among these three units (D, 
$\pi$-linker, and A) plays a vital role in facilitating efficient $\pi$ electron delocalization or photoinduced intramolecular charge transfer~\cite{charge-transfer1,charge-transfer2,charge-transfer3,planarity} from the donor to the acceptor moiety through the $\pi$-linker.~\cite{pi-linker1, pi-linker2, pi-linker3, pi-linker4, pi-linker5, pi-linker6}
 
Unfortunately, the electronic structure of a few heteroatom-doped five-membered or thiophene-type $\pi$-conjugated bridges based on a carbazole donor moiety has been studied so far and has been largely unexplored.~\cite{pi-linker4}
The lack of in-depth studies and organized data leaves a considerable gap in deepening a comprehensive understanding of how such modifications impact the optoelectronic and photophysical properties of DSSCs.
A deep exploration of these modified systems will guide us in predicting the structure-property relationships and the precise customization and design of organic materials with tailored properties for DSSCs.
To fill this crucial gap, this work presents a systematic investigation of mono-, di-, and tri-doped $\pi$-linkers, maintaining the D and A moieties as carbazole and cyanoacrylic acid, respectively, using robust and innovative pCCD-based approaches. 
Previous studies proved the reliability of these methods for electronic excited states,~\cite{eom-pccd, eom-pccd-erratum, eom-pccd-lccsd} ionization potentials,~\cite{ip-pccd, ip-tailoredpccd-jctc-2024} electron affinities,~\cite{ea-eom-pccd-jpca-2024} and transition dipole moments~\cite{lr-pccd-jctc-2024} that are comparable with the standard EOM-CCSD method. 
Utilizing pCCD-based methods enables us to surpass standard time-dependent DFT approaches in predicting electronic excited states and provide more reliable wave function calculations for such large systems in a systematic way for the first time.
\section{Methodology}
\subsection{Ground-state pCCD Model}
The pair coupled cluster doubles (pCCD) represents an efficient and cost-effective wave function method,~\cite{limacher-ap1rog-jctc-2013,pawel-pccp-geminal-review-2022,oo-ap1rog,tamar-pccd} which can be viewed as a simplified version of the conventional CCD model, where the cluster operator includes only electron-pair excitations and differs from conventional CCD approaches that incorporate  all possible double-electron excitations.
This methodology is implemented in the Pythonic Black-Box Electronic Structure Tool (PyBEST) open-source software package~\cite{pybest-paper,pybest-paper-update1-cpc-2024} to study strongly correlated closed-shell systems such as $\pi$-conjugated organic molecules.~\cite{pccd-delaram-rsc-adv-2023, ea-eom-pccd-jpca-2024} The pCCD ansatz can be written as an exponential operator acting on a reference wave function~\cite{pawel-pccp-geminal-review-2022,tamar-pccd,ap1rog-non-variational-orbital-optimization-jctc-2014}
\begin{equation} \label{eq:pccd}
\ket{\textrm{pCCD}} = e^{{T}^\textrm{pCCD}_2} \ket{\Phi_0},
\end{equation}
where ${T}_2^\textrm{pCCD}$ is a cluster operator and $\ket{\Phi_0}$ is some reference determinant (for example, Hartree--Fock)
\begin{equation}\label{eq:tp}
{T}^\textrm{pCCD}_2 
    = \sum_{i}^{n_{\textrm{occ.}}}\sum_{a}^{n_{\textrm{virt.}}} t_{i\bar{i}}^{a\bar{a}} a_{a }^{\dagger}a_{\bar{a}}^{\dagger}  a_{\bar{i}} a_{i}
    = \sum_{i}^{n_{\textrm{occ.}}}\sum_{a}^{n_{\textrm{virt.}}} t_{i\bar{i}}^{a\bar{a}} \tau_{a \bar{a} i \bar{i}}
\end{equation}
The summation in Eq.~\eqref{eq:tp} above extends over all occupied orbitals $i$ and virtual orbitals $a$, where $a_p^{\dagger}, a_{\bar{p}}^{\dagger}$, and $a_p, a_{\bar{p}}$ refer to the electron creation and annihilation operators, and $p$ and $\bar{p}$ denote spin-up ($\alpha$) and spin-down  ($\beta$) electrons, respectively, and $t_{i\bar{i}}^{a\bar{a}}$ in equation ~\eqref{eq:tp} denotes the pCCD cluster amplitudes. 
The molecular orbitals of pCCD, which serve to construct the reference determinant in Eq.~\ref{eq:pccd}, are generally optimized by applying a variational orbital optimization procedure.~\cite{oo-ap1rog, tamar-pccd, piotrus_mol-phys, ps2-ap1rog, ap1rog-non-variational-orbital-optimization-jctc-2014, pawel-pccp-geminal-review-2022}\\
A zero-valued orbital gradient of the orbital-optimized pCCD state and computed from
\begin{equation}
g_{pq} = \langle \Phi_0 | e^{-{T}_2^{\text{pCCD}}} [({E}_{pq} - {E}_{qp}), {H}] e^{{T}_2^{\text{pCCD}}} | \Phi_0 \rangle + \sum_{i,a} \lambda_i^a \left(\langle \Phi_{i\bar{i}}^{a\bar{a}} | e^{-{T}_2^{\text{pCCD}}} [({E}_{pq} - {E}_{qp}), {H}] e^{{T}_2^{\text{pCCD}}} | \Phi_0 \rangle\right)
\end{equation}
where the singlet excitation operator (${E}_{pq}$) is equal to ${p}^\dagger {q} + {\bar{p}}^{\dagger} {\bar{q}}$, with $p$ and $q$ labeling all active occupied and virtual orbitals. The Lagrange multipliers (\(\lambda_i^a\)) are computed from the pCCD \(\Lambda\) equations.~\cite{oo-ap1rog} 
\({H}\) is the molecular Hamiltonian and $\langle \Phi_{i\bar{i}}^{a\bar{a}} |$ represents a pair-excited determinant and is equal to ${a}^\dagger {\bar{a}}^\dagger  {\bar{i}}{i} | \Phi_0\rangle$.
The pCCD-optimized orbitals are localized in nature and permit the study of large ground-state electronic structures.~\cite{oo-ap1rog,pawel_jpca_2014,ap1rog-non-variational-orbital-optimization-jctc-2014,ps2-ap1rog,pawel-pccp-2015}
\subsection{pCCD Extension to Excited, Ionized, and Electron Attached States}
However, similar to the CC model, we can utilize the equation of motion (EOM) formalism~\cite{eomcc_1968,stanton-bartlett-eom-jcp-1993,bartlett-eom-cc-wires-2012} to target the electronically excited, spin-flip,~\cite{spin-flip-eom-casanova-pccp-2020} and ionized states~\cite{BN-doped,ip-tailoredpccd-jctc-2024,ip-pccd,musial2003equation} with the help of the excitation operator $\hat R$. Specifically, for excited states, we can integrate pCCD with the EOM approach that has already been proven to be successful in investigating excited states in a series of large organic systems.~\cite{domain-based-charge-transfer-lena-2025,eom-pccd,eom-pccd-erratum,eom-pccd-lccsd,pccd-delaram-rsc-adv-2023,pawel-yb2,pccd-ee-f0-actinides,eom_lcc,pccd-mocco-galynska-pccp-2024}
In the EOM-pCCD+S formalism, the $k$-th excited state wave function is expressed as
\begin{equation}
\ket{\Psi_k} = \hat R_k\ket{\textrm{pCCD}}, 
\end{equation}
where
\begin{equation}\label{eq:eom-pccd+s}
{R}_{\text{pCCD+S}} = c_0 {\tau}_0 + \sum_i^{\mathrm{occ}} \sum_a^{\mathrm{virt}} c_i^a {\tau}_{a i}+\sum_{i}^{\text{occ}} \sum_{a}^{\text{virt}} c_{i\bar{i}}^{a\bar{a}} {\tau}_{a\bar{a}i\bar{i}}
\end{equation}
The excitation operator ${R}_{\text{pCCD+S}}$ generates the target excited state through its action on the pCCD reference wave function (with or without orbital optimization). 
The first term in equation~\eqref{eq:eom-pccd+s} represents the identity operator, the second term denotes the contribution from single excitations, ${\tau}_{a i}={a}^{\dagger} {i}+{\bar{a}}^{\dagger} {\bar{i}}$, where the summation runs over all occupied and virtual orbitals. 
The last term in equation~\eqref{eq:eom-pccd+s} is the pair-excitation operator, ${\tau}_{a\bar{a}i\bar{i}} = {a}^{\dagger} {\bar{a}}^{\dagger} {\bar{i}} {i}$.
To secure electronically excited states, we diagonalize the similarity-transformed Hamiltonian of pCCD within the ${R}$ configurational space.~\cite{eom-pccd,eom-pccd-erratum,eom-pccd-lccsd}
The computational cost of the ground-state pCCD calculation is $\mathcal{O}(N^3)$.~\cite{tamar-pccd}
However, for EOM-pCCD+S, the computational scaling is $\mathcal{O}(o^2v^2)$, where $o$ and $v$ indicate the occupied and virtual orbitals, respectively, and $N = o + v$.~\cite{lr-pccd-jctc-2024}

\subsection{Calculations Details}\label{sec:calc-details}
\paragraph{Structure Optimization.}
Structural optimization was performed in a vacuum medium for mono-, di-, and tri-doped organic dyes using density functional theory (DFT) as implemented in the \textsc{Orca 6.0.0} software package,~\cite{ORCA,ORCA5} with the BP86~\cite{bp86-perdew, bp86-becke} exchange--correlational (xc) functional and the cc-pVDZ basis set~\cite{basis_dunning}.  In this initial project, our selection of the correlation-consistent basis set was made to maintain a consistent theoretical framework across different phases of our study.
Vibrational frequency calculations were carried out to ensure that all optimized structures correspond to true energy minima without imaginary frequencies. All xyz structures are provided in SI.
\paragraph{pCCD-based Calculations.}
The resulting molecular geometries were then used to investigate the electronic structure with pCCD-based methods~\cite{pawel-pccp-geminal-review-2022}, which were carried out using a developer version of the \texttt{PyBEST}~\cite{pybest-paper, pybest-paper-update1-cpc-2024} software package (\texttt{v2.1.0-dev0}). Core orbitals (1s for C, N, and O; 1s-2p for S) were kept frozen, and the cc-pVDZ basis set was used for all investigated organic dyes.
Excitation energies and CI vectors were computed using the EOM-pCCD+S method~\cite{eom-pccd,eom-pccd-erratum}.
Ionization potentials were obtained with the ionization potential equation-of-motion pCCD methods (IP-EOM-pCCD).~\cite{ip-pccd} 
The (D)IP-EOM-pCCD formalism~\cite{ip-pccd,ea-eom-pccd-jpca-2024,BN-doped} was used to compute electron affinities (EA),

\begin{align}
   E^{\rm EA} &=  E_0^{\textrm{DIP}}[3h,1p] - E_0^{\mathrm{IP}}[2h,1p] \label{eq:ea}
\end{align}

\begin{figure}[htb]
     \centering
\includegraphics[width=1.0\textwidth]{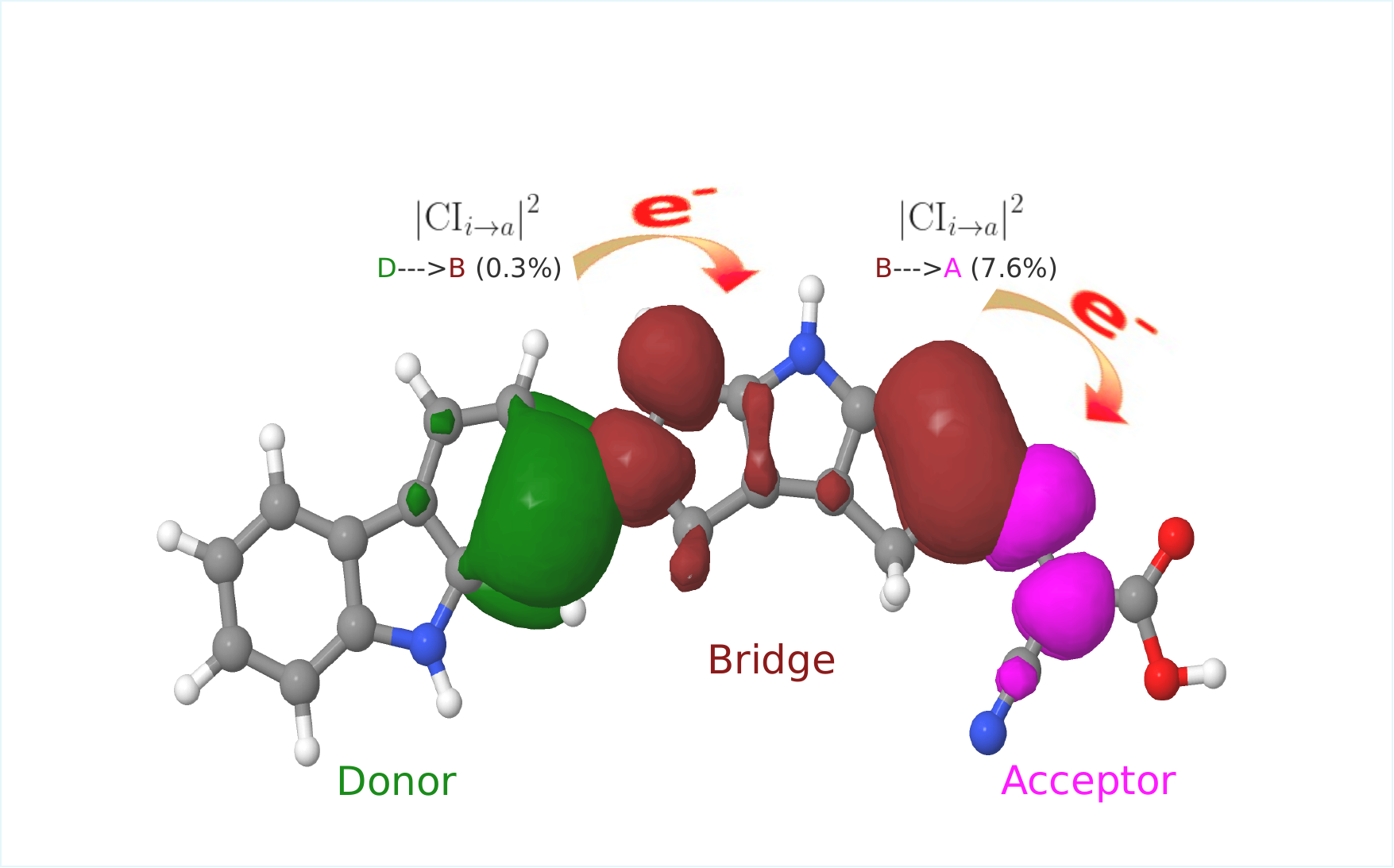}
\caption{Schematic representation of the domain-based charge-transfer analysis. All investigated dyes were divided into three domains: Donor (D), Bridge (B), and Acceptor (A).
The configuration state (CI) vectors contributing to the first excited state, each describing an individual 
$i \rightarrow a$ transition (from an occupied orbital $i$
 to an unoccupied orbital 
$a$), were analyzed and assigned to the corresponding domains. By summing the weights (squared amplitudes of the CI vectors) over each domain, the character of the excited state can be determined in terms of charge transfer between domains versus local excitation within a given domain.}
\label{fig:ct-d-b-a}
\end{figure}

Unlike standard electronic structure methods and HF canonical orbitals, pCCD natural orbitals are not strongly affected by the presence of augmented functions in the basis set. Consequently, their resulting excitation energies, ionization potentials, and electron affinities remain more stable. This observation is supported by extensive benchmarking of pCCD-based methods.~\cite{delaram-jcp,marta-jpca} The character of excited states was investigated using the recently developed domain-based charge transfer analysis~\cite{domain-based-charge-transfer-lena-2025}, the schematic representation of which is given in Figure~\ref{fig:ct-d-b-a}.
In the EOM-CC formalism, each excitation is described by a set of configuration interaction/state (CI) vectors, whose squared amplitudes provide the weights of the individual excitation components from an initial occupied orbital 
$i$ to a final unoccupied orbital $a$ within the overall state. Using these weights, together with localized pCCD orbitals, allows one to easily assign the excitation components to molecular fragments (here referred to as domains). By summing the contributions over each domain, the character of the excited state can be determined, i.e., the relative extent of charge-transfer versus local excitation within each domain. Consequently, the associated domain-specific charge transfer character can be calculated from the equation given below~\cite{domain-based-charge-transfer-lena-2025}
\begin{equation}\label{eq:domain-ct}
\%_{\mathrm{CT}(A \rightarrow B)} = \sum_{i \in A} \sum_{a \in B} |\mathrm{CI}_{i \rightarrow a}|^{2}
\end{equation}
In equation~\eqref{eq:domain-ct}, $A$ and $B$ denote two distinct molecular domains (donor (D), bridge (B), or acceptor (A)), while the symbols $i$ and $a$ are associated with their respective molecular orbitals. The term $\mathrm{CI}_{i \rightarrow a}$ represents an individual excited state vector component, where $i$ indicates an occupied orbital in domain $A$ and $a$ corresponds to a virtual orbital in domain $B$.
Such an analysis is performed automatically using a \texttt{Python} script available within the newest \texttt{PyBEST} version---\texttt{v2.2.0-dev0}.

\subsubsection{Additional DFT and TheoDORE Calculatioins}
The results obtained using pCCD-based methods were compared with trends calculated using the CAM-B3LYP~\cite{cam-b3lyp} exchange--correlation functional and the cc-pVDZ basis set, to maintain consistency with the pCCD calculations. The excited states were computed using time-dependent DFT (TD-DFT) with the same functional. The analysis of the first excited state was performed using the TheoDORE software package.~\cite{plasser2020theodore} 

\subsection{Naming Scheme for Molecular Structures}
\begin{figure}[ht!]
     \centering
\includegraphics[width=0.7\textwidth]{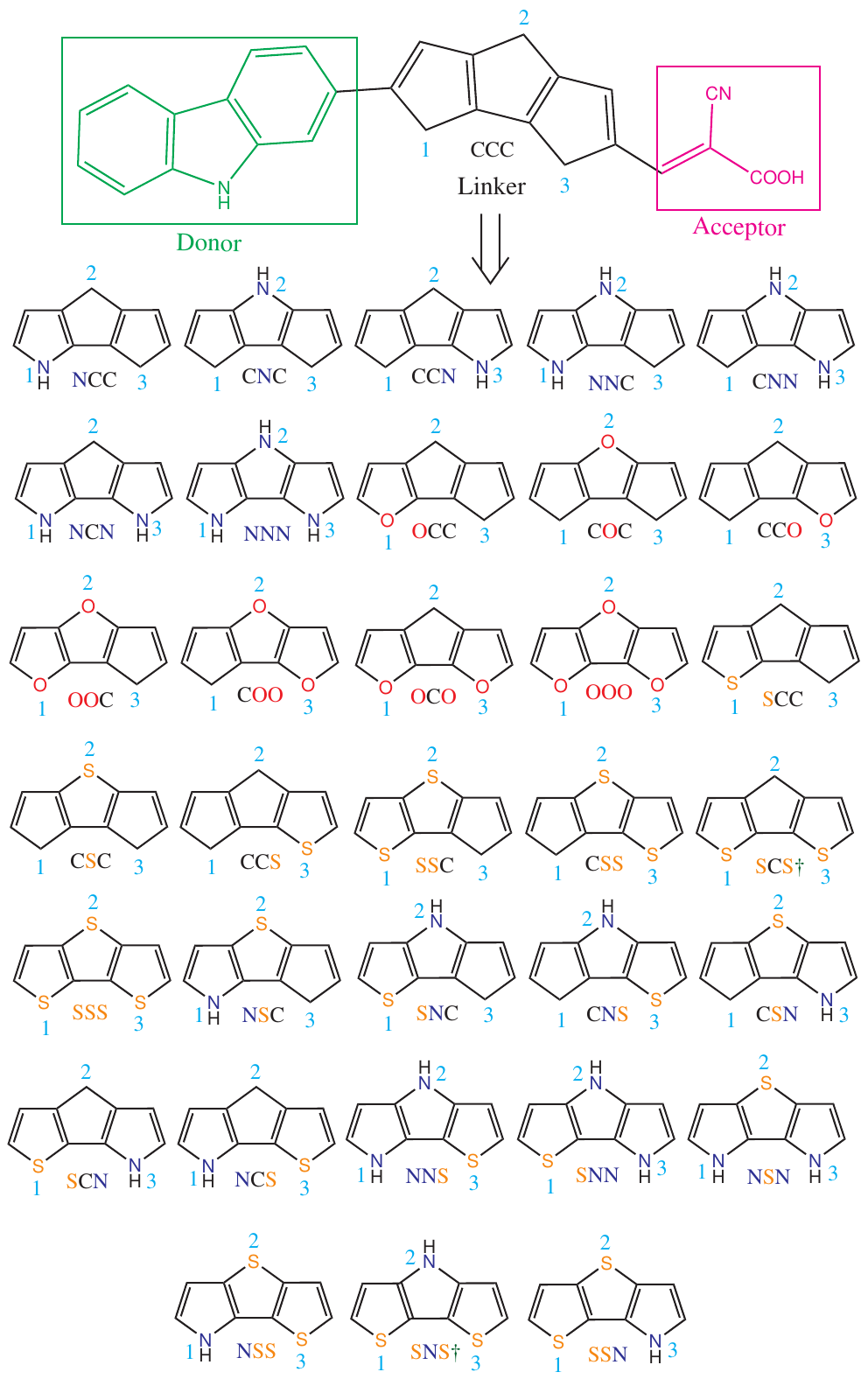}
\caption{A group of 33 $\pi$-conjugated linker moieties, mono-, di-, and tri-doped with heteroatoms ($\mathrm{{\color{nitrogen}N}}$, $\mathrm{{\color{oxygen}O}}$ and $\mathrm{{\color{sulfur}S}}$), is used to design prototypical organic dyes while maintaining a common carbazole donor and cyanoacrylic acid acceptor, where structures marked with a dagger ($\dagger$) indicate systems previously reported in the literature.~\cite{pi-linker4}}
\label{fig:doped-structures}
\end{figure}

In this work, we have designed and investigated a series of prototypical $\pi$-conjugated organic dyes by doping heteroatoms ($\mathrm{{\color{nitrogen}N}}$, $\mathrm{{\color{oxygen}O}}$ and $\mathrm{{\color{sulfur}S}}$) into the bridge of a core molecular structure (CCC) comprising a carbazole donor and a cyanoacrylic acid acceptor, connected by three fused rings of five members (referred to as the linker or bridge) and shown in the top of Figure~\ref{fig:doped-structures}.
Each ring in this bridge contains one $sp^{3}$-hybridized carbon, which does not have a lone pair of electrons to participate in conjugation with the $\pi$-bonds in the ring.
By replacing these three positions of carbon with heteroatoms, we enable their participation in $\pi$-conjugation.
This modification demonstrates the most promising strategy for extending $\pi$-conjugation and enhancing delocalization of $\pi$-electrons distribution into molecular orbitals across the entire ring, which is crucial for promoting and optimizing the charge-transfer process.~\cite{charge-transfer}
We have systematically performed mono-, di-, and tri-doping by using three distinct heteroatoms at these critical positions.
Mono- and di-doping each produces three isomers per heteroatom, yielding a total of 18 systems.
For tri-doped systems, each heteroatom corresponds to one single isomer, giving rise to 3 possible isomers.
Additionally, there are 12 mixed-type (6 di-doped and 6 tri-doped) configurations, resulting in 33 total doped organic dyes (cf. Figure~\ref{fig:doped-structures}).
The key positions are numbered as 1, 2, and 3, with a systematic naming scheme based on the heteroatoms doped at these linker positions.
For example, starting with a mono-doped system, if position 1 is doped with a nitrogen (N) atom while positions 2 and 3 remain carbon (C), the dye is named {\color{nitrogen}N}CC.
Extending this technique for a di-doped system with a nitrogen atom doped at positions 1 and 2, while position 3 is intact as carbon, the dye is labeled {\color{nitrogen}N}{\color{nitrogen}N}C. 
The naming convention of tri-doped systems follows the same rule.

\section{Results and Discussion}\label{sec:results}

\subsection{Domain-based Charge Transfer Analysis}

To demonstrate the character of the first vertical excitation of a series of 33 dyes, we present domain-based charge transfer data resolved relative to three defined molecular domains: donor (D), bridge (B), and acceptor (A), listed in Table~\ref{tab:doped_systems_final}. 
The domain-based charge transfer analysis has already been tested and validated for eight organic dyes, demonstrating how the charge transfer character influences the change in the bridge moieties.~\cite{pccd-perspective-jpcl-2023, domain-based-charge-transfer-lena-2025} 
This approach is based on the decomposition of CI vectors into nine possible contributions within these domains: D$\rightarrow$D*, D$\rightarrow$B*, D$\rightarrow$A*, B$\rightarrow$B*, B$\rightarrow$A*, *D$\leftarrow$B, A$\rightarrow$A*, *B$\leftarrow$A, and *D$\leftarrow$A. By summing the D$\rightarrow$B*, B$\rightarrow$A*, and D$\rightarrow$A* components, the forward charge transfer (D$\rightarrow$B$\rightarrow$A) is obtained, whereas considering the *B$\leftarrow$A, *D$\leftarrow$B, and *D$\leftarrow$A contributions yields the reverse charge transfer. 
The D$\rightarrow$D*, B$\rightarrow$B*, and A$\rightarrow$A* terms describe the local excitations inside each of the domains. 


\begin{table}[h!] 
\centering
\scriptsize
\setlength{\tabcolsep}{3.5pt}
\begin{tabular}{l|*{10}{c}}
\toprule
 & \multicolumn{10}{c}{\textbf{Character}} \\
\cmidrule(lr){2-11}
\textbf{System} & D$\rightarrow$D* & D$\rightarrow$B* & D$\rightarrow$A* & B$\rightarrow$B* & B$\rightarrow$A* & D*$\leftarrow$B & A$\rightarrow$A* & B*$\leftarrow$A  & D$\rightarrow$B$\rightarrow$A & D$\leftarrow$B$\leftarrow$A \\
\midrule

CCC                 & 2.7 & 4.1 & 0.4 & 63.1 & 18.2 & 2.1 & 5.0 & 1.4  & 22.8 & 3.5 \\
\multicolumn{1}{l|}{\textbf{Mono-doped}} &  &  &  &  &  &  &  &  &   \\
{\color{blue}N}CC   & 2.6 & 2.8 & 0.7 & 58.0 & 22.8 & 2.3 & 5.8 & 1.3  & 26.3 & 3.6 \\
C{\color{blue}N}C   & 3.1 & 3.4 & 0.7 & 53.3 & 25.4 & 2.6 & 6.8 & 1.3  & 29.5 & 4.0 \\
CC{\color{blue}N}   & 4.4 & 4.7 & 0.7 & 47.7 & 26.1 & 3.2 & 8.6 & 1.2  & 31.5 & 4.4 \\
{\color{red}O}CC    & 3.0 & 4.0 & 0.8 & 59.6 & 20.2 & 2.0 & 5.6 & 1.5  & 25.0 & 3.6\\
C{\color{red}O}C    & 3.4 & 4.1 & 0.7 & 56.0 & 22.2 & 2.6 & 6.2 & 1.4  & 27.0 & 4.0 \\
CC{\color{red}O}    & 4.6 & 5.4 & 0.7 & 53.4 & 20.9 & 3.0 & 7.1 & 1.5  & 27.0 & 4.5 \\
{\color{orange}S}CC & 2.8 & 3.7 & 0.5 & 61.0 & 19.6 & 1.8 & 6.0 & 1.7  & 23.8 & 3.5 \\
C{\color{orange}S}C & 3.7 & 4.9 & 0.6 & 58.9 & 19.4 & 2.3 & 5.8 & 1.6  & 25.0 & 3.8 \\
CC{\color{orange}S} & 4.0 & 5.3 & 0.7 & 57.3 & 18.9 & 2.6 & 6.2 & 1.6  & 24.9 & 4.2 \\
\midrule
\multicolumn{1}{l|}{\textbf{Di-doped}} &  &  &  &  &  &  &  &  &  &     \\
{\color{blue}N}{\color{blue}N}C  & 2.8 & 2.3 & 0.9 & 50.5 & 28.2 & 2.8 & 7.1 & 1.3 & 31.3 & 4.1 \\
C{\color{blue}N}{\color{blue}N}  & 4.4 & 3.9 & 0.9 & 40.4 & 32.1 & 3.5 & 9.7 & 1.1 & 36.9 & 4.7 \\
{\color{blue}N}C{\color{blue}N}  & 4.1 & 2.3 & 1.1 & 36.4 & 35.8 & 3.6 & 11.2 & 1.1 & 39.2 & 4.7 \\
{\color{red}O}{\color{red}O}C    & 3.7 & 4.2 & 1.0 & 53.7 & 23.2 & 2.5 & 6.6 & 1.6 &  28.4 & 4.1 \\
C{\color{red}O}{\color{red}O}    & 5.5 & 5.6 & 1.0 & 47.0 & 24.3 & 3.2 & 8.1 & 1.5 &  30.9 & 4.8 \\
{\color{red}O}C{\color{red}O}    & 5.1 & 4.8 & 1.1 & 47.2 & 25.0 & 2.9 & 8.7 & 1.7 &  31.0 & 4.6 \\
{\color{orange}S}{\color{orange}S}C & 3.8 & 4.2 & 0.6 & 56.8 & 20.6 & 2.0 & 6.8 & 1.8  & 25.4 & 3.8 \\
C{\color{orange}S}{\color{orange}S} & 5.5 & 6.6 & 0.8 & 55.2 & 18.3 & 2.7 & 6.2 & 1.6  & 25.7 & 4.3 \\
{\color{orange}S}C{\color{orange}S} & 3.8 & 4.6 & 0.6 & 56.7 & 20.4 & 2.0 & 7.1 & 1.8  & 25.6 & 3.8 \\
{\color{blue}N}{\color{orange}S}C   & 3.3 & 3.0 & 0.8 & 54.6 & 23.4 & 2.6 & 6.7 & 1.6  & 27.1 & 4.2 \\
C{\color{blue}N}{\color{orange}S}   & 4.6 & 4.6 & 0.8 & 50.5 & 24.1 & 3.1 & 7.4 & 1.4 & 29.5 & 4.6 \\
{\color{blue}N}C{\color{orange}S}   & 3.4 & 3.2 & 0.9 & 53.7 & 24.0 & 2.6 & 6.9 & 1.4  & 28.1 & 4.0 \\
{\color{orange}S}{\color{blue}N}C   & 2.9 & 3.0 & 0.7 & 52.8 & 26.0 & 2.0 & 7.5 & 1.6  & 29.7 & 3.6 \\
C{\color{orange}S}{\color{blue}N}   & 5.3 & 5.3 & 0.9 & 44.5 & 27.3 & 3.0 & 9.1 & 1.2  & 33.6 & 4.3 \\
{\color{orange}S}C{\color{blue}N}   & 4.1 & 3.7 & 0.7 & 43.1 & 30.0 & 2.6 & 10.7 & 1.5  & 34.4 & 4.1 \\

\midrule
\multicolumn{1}{l|}{\textbf{Tri-doped}} &  &  &  &  &    &  &  &  &  &  \\
{\color{oxygen}O}{\color{oxygen}O}{\color{oxygen}O} & 6.1 & 5.3 & 1.5 & 42.4 & 27.2 & 3.2 & 9.4 & 1.6 & 34.0 & 4.8 \\
{\color{orange}S}{\color{orange}S}{\color{orange}S} & 5.5 & 5.4 & 0.8 & 52.8 & 19.6 & 2.4 & 7.5 & 2.0 & 25.8 & 3.5 \\
{\color{blue}N}{\color{orange}S}{\color{orange}S}   & 4.6 & 3.7 & 1.0 & 50.1 & 24.0 & 2.9 & 7.8 & 1.6 & 28.8 & 4.6 \\
{\color{orange}S}{\color{blue}N}{\color{orange}S}   & 4.8 & 3.9 & 0.8 & 48.8 & 25.0 & 2.6 & 8.4 & 1.7 & 29.6 & 4.3 \\
{\color{orange}S}{\color{orange}S}{\color{blue}N}   & 4.8 & 4.1 & 0.9 & 40.6 & 30.7 & 2.4 & 11.1 & 1.5 & 35.8 & 3.9 \\
{\color{blue}N}{\color{blue}N}{\color{orange}S}     & 4.0 & 2.8 & 1.0 & 46.5 & 28.4 & 3.3 & 8.3 & 1.5 & 32.3 & 4.8 \\
{\color{orange}S}{\color{blue}N}{\color{blue}N}     & 4.4 & 3.0 & 0.9 & 36.9 & 34.8 & 3.0 & 11.5 & 1.3 & 38.6 & 4.3 \\
{\color{blue}N}{\color{orange}S}{\color{blue}N}     & 4.4 & 2.4 & 1.2 & 34.6 & 36.7 & 3.2 & 11.9 & 1.2 & 40.3 & 4.4 \\
{\color{blue}N}{\color{blue}N}{\color{blue}N}       & 4.0 & 2.0 & 1.2 & 32.2 & 39.5 & 3.8 & 11.7 & 1.0 & 42.6 & 4.8 \\
\bottomrule
\end{tabular}
\begin{tablenotes}\footnotesize
\item The asterisk($^\star$) denotes excitation to an unoccupied (virtual) orbital in different domains relative to the pCCD reference determinant.
\end{tablenotes}
\caption{The domain-based charge transfer percentages (\% CT) contributing to the first excited state of mono-, di-, and tri-doped systems calculated using the EOM-pCCD+S method and the cc-pVDZ basis set.
The transitions involved from occupied to virtual orbitals across different domain orbitals, with both forward (D$\rightarrow$B$\rightarrow$A) and reverse (D$\leftarrow$B$\leftarrow$A) total charge transfer.
The D*$\leftarrow$A contribution is not shown as it is below 0.1\% for all investigated systems.}
\label{tab:doped_systems_final}
\end{table}

\subsubsection{Dyes with Mono-doped Bridges}
In mono-doped systems, the donor-to-bridge (D$\rightarrow$B*) charge transfer percentage increases as the doping position shifts closer to the acceptor domain.
This trend is observed in all mono-doped systems: nitrogen ($\mathrm{{\color{nitrogen}N}CC}\textless\mathrm{C{\color{nitrogen}N}C}\textless\mathrm{CC{\color{nitrogen}N}}$), oxygen ($\mathrm{{\color{oxygen}O}CC}\textless\mathrm{C{\color{oxygen}O}C}\textless\mathrm{CC{\color{oxygen}O}}$), and sulfur ($\mathrm{{\color{sulfur}S}CC}\textless\mathrm{C{\color{sulfur}S}C}\textless\mathrm{CC{\color{sulfur}S}}$) with exact values listed in Table~\ref{tab:doped_systems_final}.
Two primary factors are responsible for this behavior. First, the observed enhancement results from improved planarity between the donor and bridge moieties as the dihedral angle decreases, which allows more efficient $\pi$-electron delocalization.~\cite{dihedral-angle} This effect arises from the heteroatom's migration from either position 1 to position 2 or 3.  However, oxygen's small size as a dopant minimizes steric hindrance between donor and bridge moieties~\cite{steric-hindrance}, preventing significant alteration of this angle, as shown in Figure S1 and Table S1 in the SI. Second, the electron density of the bridge shifts slightly as the heteroatoms move from position 1 towards position 3. This sequential relocation induces a gradual change in electron density distribution. As a result, the left side of the bridge near the donor becomes slightly electron-deficient, facilitating electron acceptance from the donor moiety. The bridge-to-acceptor (B$\rightarrow$A*) charge transfer exhibits distinct trends for each dopant, increasing for nitrogen, decreasing for sulfur, and displaying non-straightforward behavior for oxygen-doped systems.  The dihedral angle between the $\pi$-bridge and acceptor approaches zero in all systems, indicating near-perfect planarity, as depicted in Figure S2 and Table S2 in the SI. Therefore, this charge transfer component depends on two key factors:
(1) heteroatom electronegativity (governing lone pair binding strength and thus donation capacity), (2)  spatial compatibility between $\mathrm{{\color{nitrogen}N}}$/ $\mathrm{{\color{oxygen}O}}$/ $\mathrm{{\color{sulfur}S}}$ lone pair orbitals and acceptor $\pi$ orbitals, which facilitates effective overlap and delocalization of $\pi$-electron density. The overall forward charge transfer ($\mathrm{D}\rightarrow\mathrm{B}\rightarrow\mathrm{A}$) consistently increases as the position of heteroatoms changes from 1 to 2 to 3 ($\mathrm{{\color{nitrogen}N}CC (26.3
\%)}\textless\mathrm{C{\color{nitrogen}N}C (29.5\%)}\textless\mathrm{CC{\color{nitrogen}N} (31.5\%)}$, $\mathrm{{\color{oxygen}O}CC (24.96\%)}\textless\mathrm{C{\color{oxygen}O}C (27.0\%)}=\mathrm{CC{\color{oxygen}O}(27.0\%)}$, and $\mathrm{{\color{sulfur}S}CC (23.8\%)}\textless\mathrm{CC{\color{sulfur}S} (24.9\%)}=\mathrm{C{\color{sulfur}S}C}$ (25.0\%), as shown in Table~\ref{tab:doped_systems_final} and Figure~\ref{fig:contribution}. The reverse  charge transfer (D$\leftarrow$B$\leftarrow$A) is in the range of 3.5\% to 4.5\% for all systems, which is significantly lower than the forward charge transfer, as given in Table~\ref{tab:doped_systems_final}. The nitrogen-doped system shows higher forward charge transfer characteristics in comparison to both oxygen- and sulfur-doped systems. These results suggest that mono-doping of nitrogen at position~3 ($\mathrm{CC{\color{nitrogen}N}}$) emerges as the optimal site, exhibiting the highest (31.5\%) forward charge transfer efficiency (see Table~\ref{tab:doped_systems_final} and Figure~\ref{fig:contribution}).

\subsubsection{Dyes with Di-doped Bridges}

Di-doped systems can be classified into two types based on the dopant pair: homogeneous (two identical heteroatoms, $\mathrm{{\color{nitrogen}N}}$  or $\mathrm{{\color{sulfur}S}}$ or $\mathrm{{\color{oxygen}O}}$) and heterogeneous (two different heteroatoms, $\mathrm{{\color{nitrogen}N}}$ and $\mathrm{{\color{sulfur}S}}$), as depicted in Table~\ref{tab:doped_systems_final} and Figure~\ref{fig:doped-structures}.
In all homogeneous and heterogeneous di-doped systems, the D$\rightarrow$B* charge transfer value follows the same trend.
When examining the series of three dyes in which the same atoms exchange doping positions, the highest D$\rightarrow$B* values are observed for systems with position 1 undoped, whereas the smallest values occur when position 3 is undoped. This suggests that the dihedral angle between the donor and the bridge plays a significant role, as all dyes with carbon in position 1 exhibit angles close to 0$^\circ$. A similar behavior is also observed for oxygen-doped dyes, which show noticeable D$\rightarrow$B* values and dihedral angles close to 0$^\circ$.
Similarly to D$\rightarrow$B*, the B$\rightarrow$A* charge transfer is the smallest (except for homogeneous sulfur-doped dyes), when position 3 is undoped. 

The forward (D$\rightarrow$B$\rightarrow$A) charge transfer increases progressively for  the $\mathrm{{\color{nitrogen}N}}$- and $\mathrm{{\color{oxygen}O}}$-doped dyes: 
$\mathrm{{\color{nitrogen}N}{\color{nitrogen}N}C (31.3\%)}\textless\mathrm{C{\color{nitrogen}N}{\color{nitrogen}N }(36.9\%)}\textless\mathrm{{\color{nitrogen}N}C{\color{nitrogen}N}(39.2\%)}$ and $\mathrm{{\color{oxygen}O}{\color{oxygen}O}C (28.4\%)}\textless\mathrm{C{\color{oxygen}O}{\color{oxygen}O} (30.9\%)}\textless\mathrm{{\color{oxygen}O}C{\color{oxygen}O} (31.0\%)}$. 
In case of sulfur doping the D$\rightarrow$B$\rightarrow$A values are almost the same: $\mathrm{{\color{sulfur}S}{\color{sulfur}S}C (25.4\%)}\approx\mathrm{{\color{sulfur}S}C{\color{sulfur}S}(25.6\%)}\approx
\mathrm{C{\color{sulfur}S}{\color{sulfur}S }(25.7\%)}$. In general, the homogeneous di-doped systems show the smallest forward charge transfer when position 3 is undoped. 

In mixed systems, the trend for the forward charge transfer is $\mathrm{{\color{nitrogen}N\color{sulfur}S}C}\textless\mathrm{{\color{nitrogen}N}C\color{sulfur}S}\textless\mathrm{{C\color{nitrogen}N\color{sulfur}S}}\textless\mathrm{{\color{sulfur}S}{\color{nitrogen}N}C}\textless \mathrm{C{\color{sulfur}S}{\color{nitrogen}N}}\textless\mathrm{{\color{sulfur}S}C{\color{nitrogen}N}}$. 
The charge transfer increases gradually as the nitrogen atom is positioned closer to the acceptor moiety (1$\rightarrow$2$\rightarrow$3). This enables nitrogen's lone pairs (a stronger donor than sulfur) to enhance conjugation and electron delocalization within the $\pi$-system.
The charge transfer in nitrogen-sulfur (mixed) di-doped systems is higher than that in pure sulfur-doped systems but lower than in pure nitrogen-doped systems. These results indicate that di-doped nitrogen reveals higher charge transfer in comparison to oxygen- and sulfur-doped analogs.

\subsubsection{Dyes with Tri-doped Bridges}
Tri-doped systems can also be categorized into pure and mixed types. Pure systems contain three identical dopants, such as $\mathrm{{\color{oxygen}O}{\color{oxygen}O}{\color{oxygen}O}}$, $\mathrm{{\color{sulfur}S}{\color{sulfur}S}{\color{sulfur}S}}$, and $\mathrm{{\color{nitrogen}N}{\color{nitrogen}N}{\color{nitrogen}N}}$ (presented in Table~\ref{tab:doped_systems_final}). The second, mixed type, comprises systems containing either one nitrogen and two sulfur atom or two nitrogen and one sulfur atoms as illustrated in Table~\ref{tab:doped_systems_final} and Figure~\ref{fig:doped-structures}. In pure doping, D$\rightarrow$B* increases while B$\rightarrow$A* decreases as the heteroatom changes from nitrogen to oxygen to sulfur. As a result, D$\rightarrow$B$\rightarrow$A decreases with changing dopant from nitrogen (42.6\%) to oxygen (34.0\%) to sulfur (25.8\%). 

The first mixed-doping type includes one nitrogen and two sulfur atoms, where  D$\rightarrow$B* and B$\rightarrow$A* increase upon changing the position of the nitrogen heteroatom from 1 to 2 to 3 ($\mathrm{{\color{nitrogen}N}{\color{sulfur}S}{\color{sulfur}S}}\rightarrow\mathrm{{\color{sulfur}S}{\color{nitrogen}N}{\color{sulfur}S}}\rightarrow\mathrm{{\color{sulfur}S}{\color{sulfur}S}{\color{nitrogen}N}}$) in the bridge as depicted in Table~\ref{tab:doped_systems_final} and Figure~\ref{fig:doped-structures}. Consequently, D$\rightarrow$B$\rightarrow$A increases from 28.8\% to 29.6\% to 35.8\%, and D$\leftarrow$B$\leftarrow$A decreases from 4.6\% to 4.3\% to 3.9\%.  
The second mixed-doping type comprises systems with two nitrogen and one sulfur atom. In this series, the D$\rightarrow$B* contribution increase in a series 
$\mathrm{{\color{nitrogen}N}{\color{sulfur}S}{\color{nitrogen}N}}\textless\mathrm{{\color{nitrogen}N}{\color{nitrogen}N}{\color{sulfur}S}}\textless\mathrm{{\color{sulfur}S}{\color{nitrogen}N}{\color{nitrogen}N}}$.
This trend indicates that position 1 plays a critical role: when this site is doped with a nitrogen atom, the D$\rightarrow$B* charge transfer decreases, likely due to the depletion of electron density on the bridge caused by the electron-donating character of nitrogen.
The B$\rightarrow$A* and D$\rightarrow$B$\rightarrow$A
increases with the change in the position of nitrogens from 1 and 2 to 2 and 3 to 1 and 3. 

\subsubsection{Summary of Doping Strategies }
Collectively, our results demonstrate that modifying the bridge region of the core CCC framework through mono-, di-, or tri-doping with heteroatoms substantially enhances charge transfer, yielding average increases of up to 3.8\%, 7.6\%, and 11.4\%, respectively, as shown in Table~\ref{tab:doped_systems_final}.
The highest charge transfer is observed for dyes doped with nitrogen in each series (mono-doped: $\mathrm{CC{\color{nitrogen}N}}$; di-doped: $\mathrm{{\color{nitrogen}N}C{\color{nitrogen}N}}$; tri-doped: $\mathrm{{\color{nitrogen}N}{\color{nitrogen}N}{\color{nitrogen}N}}$), and it increases with the degree of doping.
In $\mathrm{{\color{nitrogen}N}C{\color{nitrogen}N}}$, the D$\rightarrow$B* charge transfer decreases by 2.4\%, while the B$\rightarrow$A* component increases by 9.7\%, resulting in an overall 7.7\% increase in forward charge transfer compared to the mono-doped $\mathrm{CC{\color{nitrogen}N}}$ dye.
For $\mathrm{{\color{nitrogen}N}{\color{nitrogen}N}{\color{nitrogen}N}}$, the same trend is maintained: the D$\rightarrow$B* contribution decreases further by 0.3\%, while the B$\rightarrow$A* part increases by 3.7\%. Interestingly, increasing the degree of doping shifts increase coupling between bridge and acceptor, rather than reinforcing it on the donor domain. 

\begin{figure}
    \centering
    \includegraphics[width=1.0\linewidth]{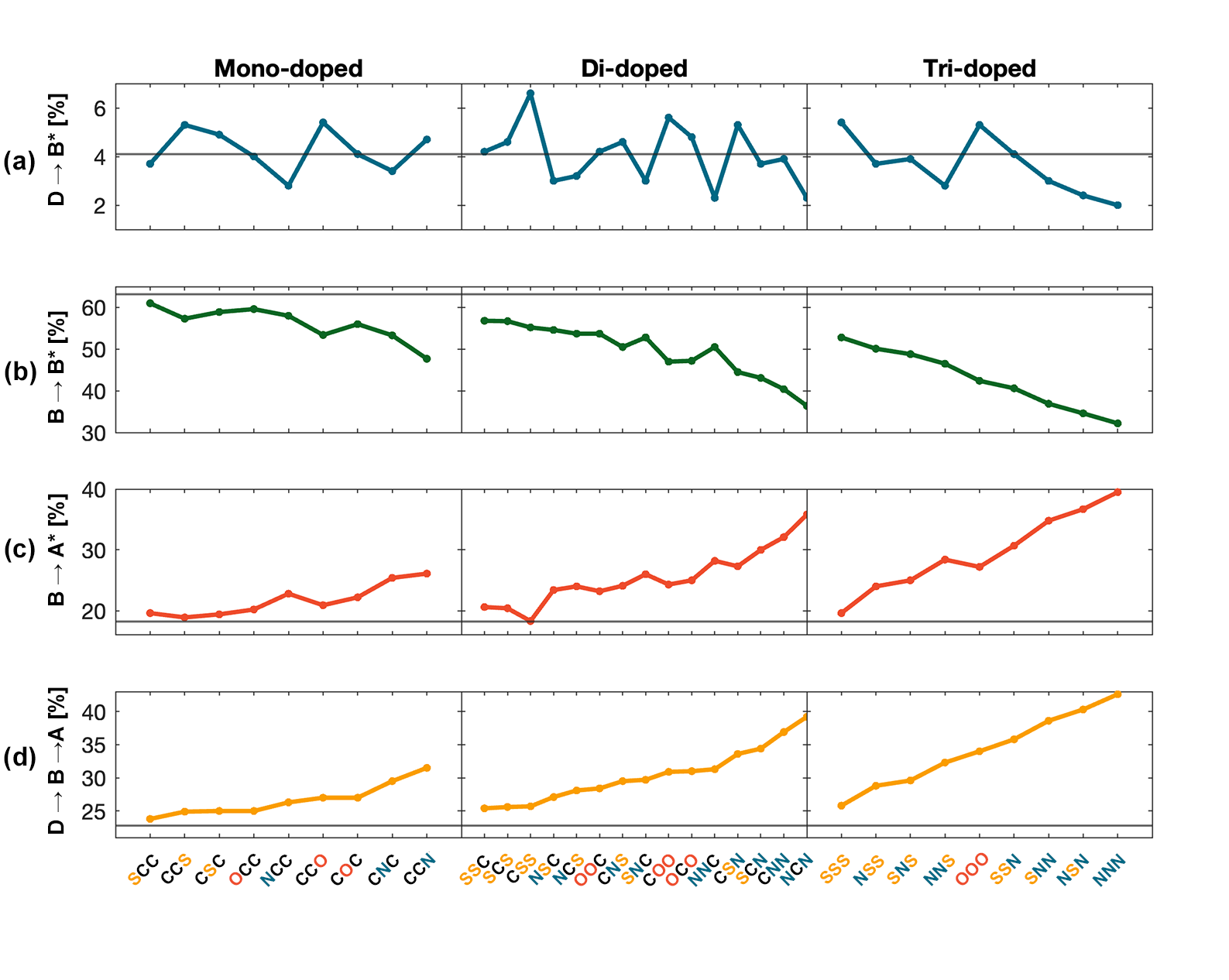}
    \caption{The (a) D$\rightarrow$B*, (b) B$\rightarrow$B*, (c) B$\rightarrow$A*, and (d) D$\rightarrow$B$\rightarrow$A contributions to the first excited state of the mono-, di-, and tri-doped systems, calculated using the EOM-pCCD+S/cc-pVDZ method. The data was sorted according to the forward charge transfer.}
    \label{fig:contribution}
\end{figure}

The D$\rightarrow$B*, B$\rightarrow$B*,  B$\rightarrow$A*, and overall D$\rightarrow$B$\rightarrow$A charge transfer components are plotted in Figure~\ref{fig:contribution} in order of increased forward charge transfer. In general, the dominant contribution to charge transfer goes from the B$\rightarrow$A* pathway, accounting for approximately 20–40\% of the total. This contribution increases with the number of doped sites in the bridge. In contrary, the largest D$\rightarrow$B* contributions are independent on the number of doped sites and are mostly observed in systems with homogeneous sulfur or oxygen doping.
However, combining sulfur and nitrogen in mixed-doped systems does not yield higher charge transfer than that observed for $\mathrm{{\color{nitrogen}N}{\color{nitrogen}N}{\color{nitrogen}N}}$. On the other hand, the performance of oxygen-doped systems is significantly better than that of sulfur-doped ones, suggesting that mixed oxygen–nitrogen doping could promote greater charge separation and higher charge transfer compared to sulfur–nitrogen combinations.

Since the main excitation occurs on the bridge, doping allows tuning of the ratio between B$\rightarrow$A* and B$\rightarrow$B* contributions relative to the undoped (CCC) system. In CCC, the B$\rightarrow$A* value is the smallest, while B$\rightarrow$B* is the largest among all investigated dyes. 
Upon doping, the B$\rightarrow$B* values (Figure~\ref{fig:contribution}(b)) appear to vary inversely with the B$\rightarrow$A* contributions (Figure~\ref{fig:contribution}(c)).


\subsubsection{Hole and Particle Character Across the Domains in EOM-pCCD+S and TD-CAM-B3LYP}

\begin{figure}
    \centering
    \includegraphics[width=1.0\linewidth]{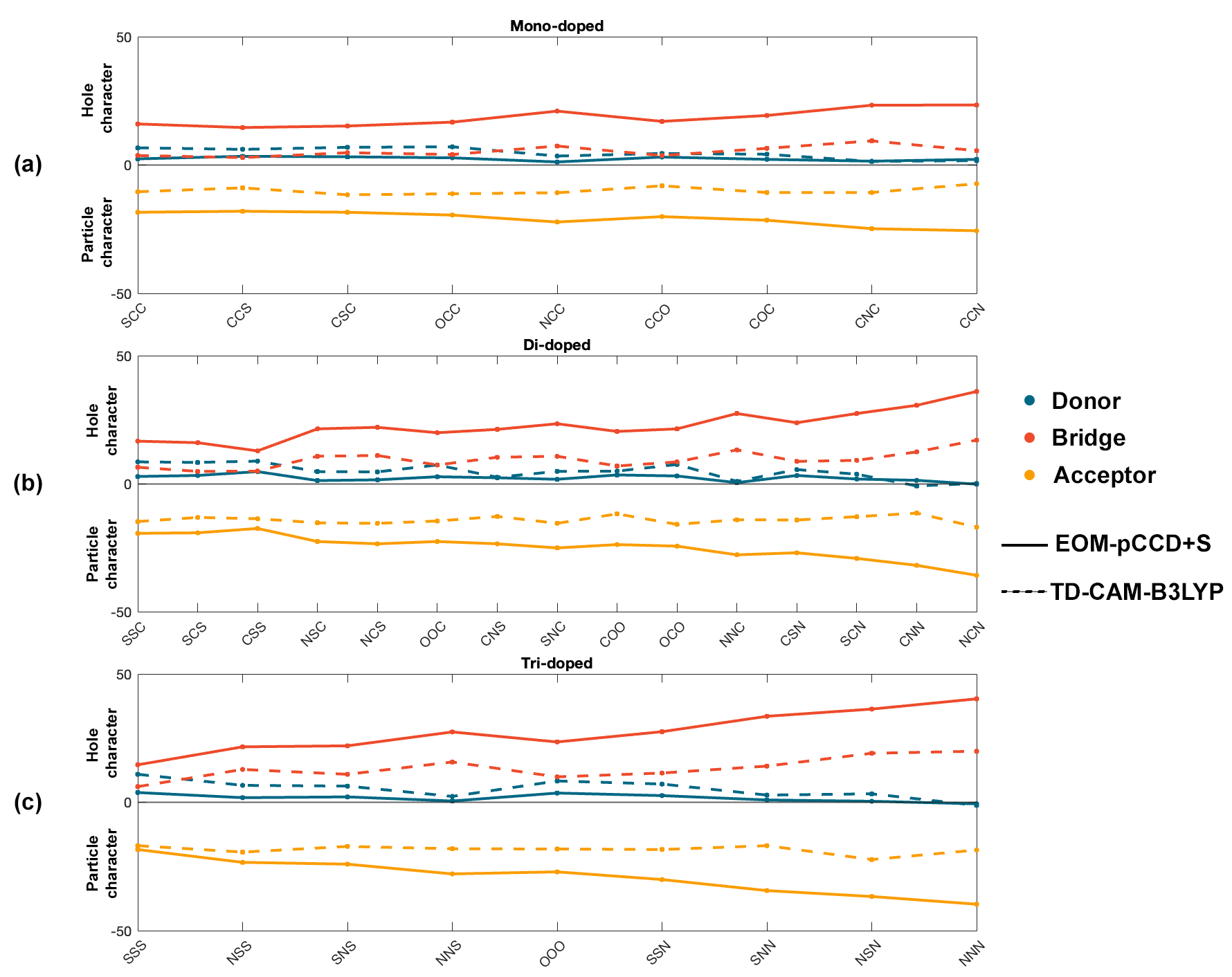}
    \caption{The hole and electron character for each domain (donor, bridge, and acceptor marked in blue, red, and yellow, respectively), obtained by subtracting the hole population from the electron population from the CAM-B3LYP (dash line) and EOM-pCCD+S calculations obtained for dyes with (a) mono-, (b) di-, and (c) tri-doped bridges. Positive values indicate a greater hole character, whereas negative values suggest a greater electron character. }
    \label{fig:hole-particle}
\end{figure}

The results obtained from the domain-based decomposition analysis were cross-checked and compared with data calculated using TD-CAM-B3LYP and further examined through transition density matrix analysis using TheoDORE\cite{plasser2020theodore}. 
Since these are entirely different methodologies, the previously calculated charge transfer values cannot be directly compared. 
Therefore, using both analyses, we determined the dominant hole and particle character for each domain, with the results shown in Figure~\ref{fig:hole-particle}. 
The hole (particle) characters from the domain-based decomposition analysis were calculated by summing the outgoing (incoming) contributions to each domain, including local excitations. 
For example, to calculate the hole character in the donor, we sum D$\rightarrow$B*, D$\rightarrow$A*, and D$\rightarrow$D*, while to obtain the particle character, we sum *D$\leftarrow$B, *D$\leftarrow$A, and D$\rightarrow$D*.
To determine the dominant character on each domain, the hole and particle characters were subtracted from each other, with positive values indicating a predominance of hole character and negative values indicating a predominance of particle character on the domain. 

The EOM-pCCD+S hole and particle character exhibit significantly larger values on the bridge and acceptor domains than those of TD-CAM-B3LYP. This observation is in line what we observed previously---EOM-pCCD+S tends to overestimate excitation energies in $\pi$-conjugated systems, such as all trans-polyene systems,~\cite{trans-polyenes} and actinide compounds.~\cite{actinide} In contrast, TD-DFT can also fail. For instance, the TD-DFT/CAM-B3LYP functional is known to underestimate excitation energies in charge-transfer states drastically.~\cite{TDDFT-fails}
Even with these fundamental methodological differences, the overall trends from both approaches are very comparable (cf. Figure~\ref{fig:hole-particle}). 
However, based on Figure~\ref{fig:hole-particle} some differences are noticeable. For instance, in the homogeneously sulfur-doped dyes, the donor contributes more to the charge transfer than the bridge, suggesting slightly larger charge separation in these systems when evaluated with CAM-B3LYP. This behavior is also reflected in the larger D$\rightarrow$B* values obtained with EOM-pCCD+S, although in this method the hole remains primarily localized on the bridge. Nevertheless, both approaches predict relatively weak overall charge separation, with the charge transfer occurring predominantly from the bridge to the acceptor.

\subsection{Electronic Properties}
\begin{table}[ht!]
\centering
\scriptsize
\begin{tabular}{c|c|c|c|c}
\toprule
Doped-system & IP [eV] & EA [eV] & IP-EA [eV] & FEE [eV]\\
\midrule
CCC         & 4.22 & -0.07 & 4.29 & 3.68 \\
\midrule
{\color{nitrogen}N}CC                                     & 4.01 & 0.25 & 3.76 & 3.70\\
C{\color{nitrogen}N}C                                     & 4.01 & 0.20 & 3.81 & 3.75\\
CC{\color{nitrogen}N}                                     & 4.20 & 0.28 & 3.92 & 3.94\\
{\color{nitrogen}N}{\color{nitrogen}N}C                   & 3.90 & 0.49 & 3.41 & 3.75\\
C{\color{nitrogen}N}{\color{nitrogen}N}                   & 4.01 & 0.42 & 3.59 & 3.92\\
{\color{nitrogen}N}C{\color{nitrogen}N}                   & 4.01 & 0.81 & 3.20 & 3.98 \\
{\color{nitrogen}N}{\color{nitrogen}N}{\color{nitrogen}N} & 3.92 & 0.88 & 3.04 & 3.97 \\
\hline
{\color{oxygen}O}CC                                       & $\star$ & $\star$ & $\star$ & 3.77\\
C{\color{oxygen}O}C                                       & 4.30 & -0.04 & 4.34 & 3.77 \\
CC{\color{oxygen}O}                                       & 4.36 &  0.14 & 4.22 & 3.91 \\
{\color{oxygen}O}{\color{oxygen}O}C                       & 4.40 &  0.08 & 4.32 & 3.85\\
C{\color{oxygen}O}{\color{oxygen}O}                       & 4.47 &  0.14 & 4.33 & 3.96\\
{\color{oxygen}O}C{\color{oxygen}O}                       & 4.44 &  0.34 & 4.10 & 4.05\\
{\color{oxygen}O}{\color{oxygen}O}{\color{oxygen}O}       & 4.57 &  0.21 & 4.36 & 4.08 \\
\hline
{\color{sulfur}S}CC                                       & 4.41 & -0.01 & 4.42 & 3.90 \\
C{\color{sulfur}S}C                                       & 4.40 & -0.04 & 4.44 & 3.93 \\
CC{\color{sulfur}S}                                       & 4.39 &  0.02 & 4.37 & 3.88 \\
{\color{sulfur}S}{\color{sulfur}S}C                       & 4.58 &  0.03 & 4.55 & 4.11 \\
C{\color{sulfur}S}{\color{sulfur}S}                       & 4.56 & -0.03 & 4.59 & 4.12 \\
{\color{sulfur}S}C{\color{sulfur}S}                       & 4.57 &  0.12 & 4.45 & 4.11 \\
{\color{sulfur}S}{\color{sulfur}S}{\color{sulfur}S}       & 4.78 &  0.10 & 4.68 & 4.29 \\
\hline
{\color{nitrogen}N}{\color{sulfur}S}C                     & 4.32 & 0.24 & 4.08 & 3.96 \\
{\color{sulfur}S}{\color{nitrogen}N}C                     & 4.25 & 0.25 & 4.00 & 3.95 \\
C{\color{nitrogen}N}{\color{sulfur}S}                     & 4.24 & 0.18 & 4.06 & 3.99 \\
C{\color{sulfur}S}{\color{nitrogen}N}                     & 4.38 & 0.24 & 4.14 & 4.09 \\
{\color{sulfur}S}C{\color{nitrogen}N}                     & 4.38 & 0.43 & 3.95 & 4.12 \\
{\color{nitrogen}N}C{\color{sulfur}S}                     & 4.19 & 0.34 & 3.85 & 3.92 \\
{\color{nitrogen}N}{\color{nitrogen}N}{\color{sulfur}S}   & 4.12 & 0.52 & 3.60 & 4.00 \\
{\color{sulfur}S}{\color{nitrogen}N}{\color{nitrogen}N}   & 4.26 & 0.56 & 3.70 & 4.09 \\
{\color{nitrogen}N}{\color{sulfur}S}{\color{nitrogen}N}   & 4.25 & 0.73 & 3.52 & 4.14 \\
{\color{sulfur}S}{\color{sulfur}S}{\color{nitrogen}N}     & 4.56 & 0.43 & 4.13 & 4.24 \\
{\color{nitrogen}N}{\color{sulfur}S}{\color{sulfur}S}     & 4.44 & 0.34 & 4.10 & 4.15 \\
{\color{sulfur}S}{\color{nitrogen}N}{\color{sulfur}S}     & 4.49 & 0.27 & 4.22 & 4.18 \\
\bottomrule
\end{tabular}
\begin{tablenotes}\footnotesize
\centering
\item The asterisk($^\star$) indicates convergence issues related to Hartree-Fock.
\end{tablenotes}
\caption{Ionization potentials (IPs), electron affinities (EAs), HOMO-LUMO gaps (IP-EA), and first excitation energies (FEE) obtained from the pCCD-based methods with the cc-pVDZ basis set.}
\label{tab:IP_results}
\end{table}

\begin{figure}
    \centering
    \includegraphics[width=1.0\linewidth]{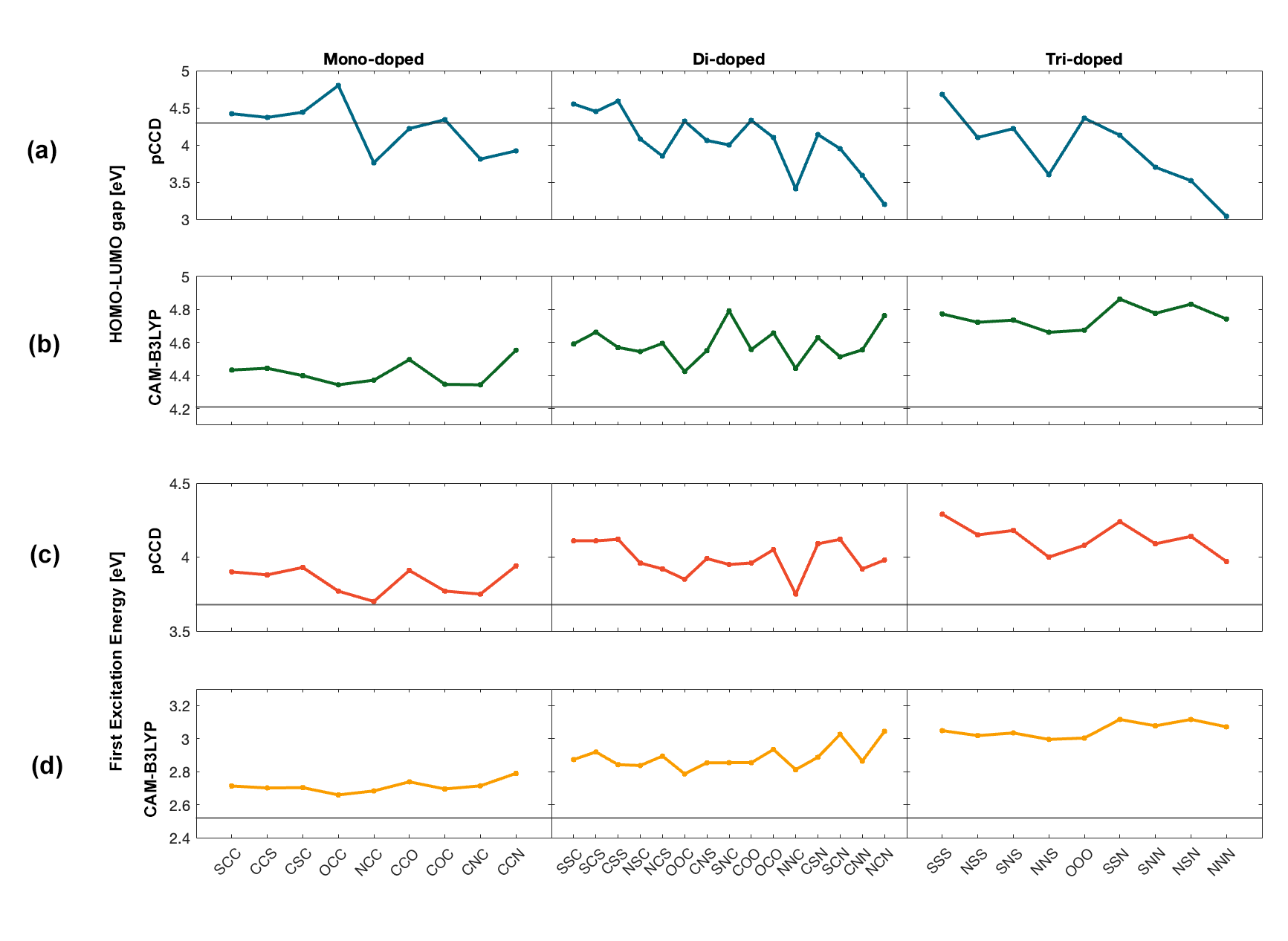}
    \caption{(a,b) HOMO–LUMO gaps and (c,d) first excitation energies calculated with (a, c) EOM-pCCD+S/cc-pVDZ and (b, d) TD-CAM-B3LYP/cc-pVDZ methods. The gray line is the reference value calculated for the undoped (CCC) system.}
    \label{fig:HL_FEE}
\end{figure}

\begin{figure}
    \centering
    \includegraphics[width=1.0\linewidth]{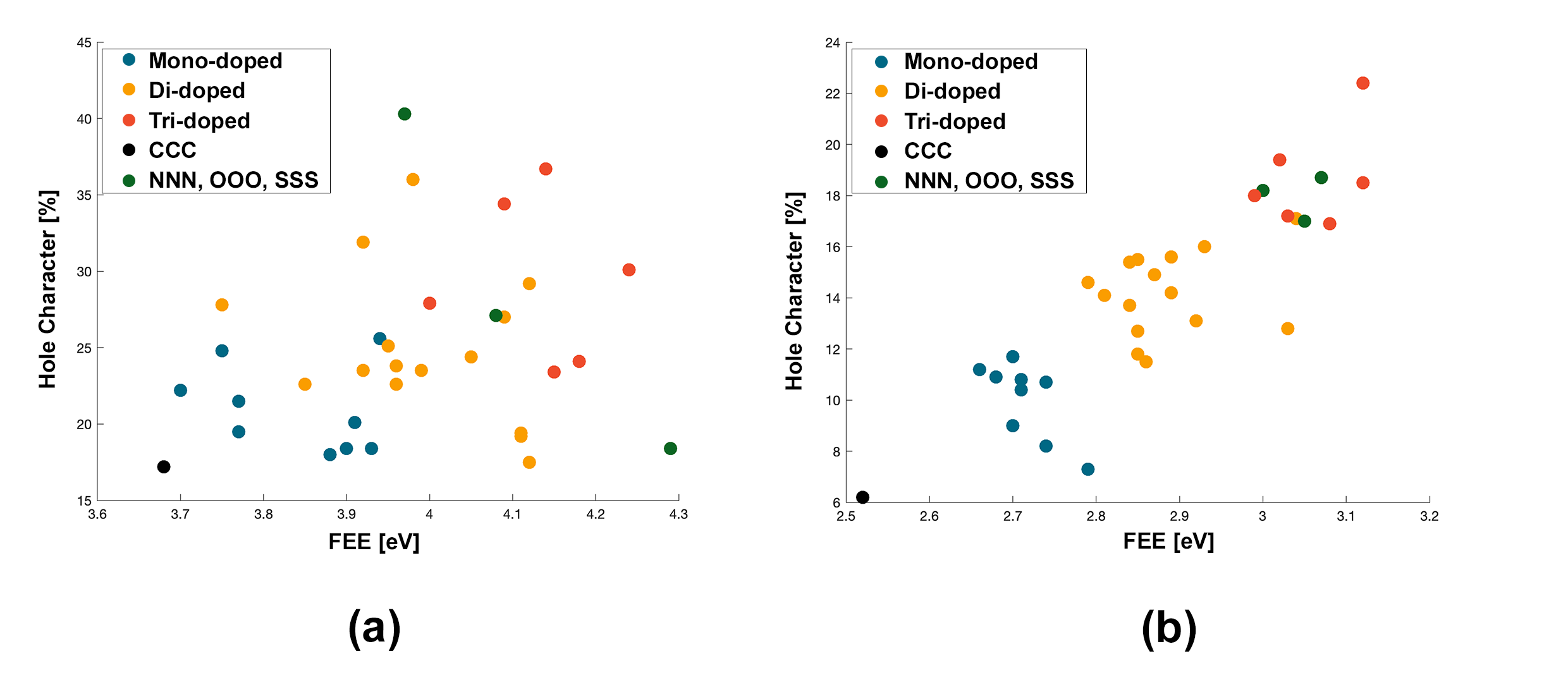}
    \caption{Dependence of the first excitation energy on the total hole character calculated with (a) EOM-pCCD+S and (b) TD-CAM-B3LYP.}
    \label{fig:hole_FEE}
\end{figure}

Finally, we will scrutinize if the calculated charge transfer trends correlate somehow with the electronic properties predicted with pCCD-based and DFT-based methods.
To that end, we invesigated the HOMO-LUMO gaps calculated from ionization potentials (IPs) and electron affinities (EAs) in case of the pCCD-based methods and orbital energies in case of CAM-B3LYP, as well as the first excitation energies (FEEs) calculated from EOM-pCCD+S and TD-CAM-B3LYP as shown in Figure~\ref{fig:HL_FEE} with the same ordering of the dyes as used in Figure~\ref{fig:contribution}, while  Figure~\ref{fig:hole_FEE} shows the dependence on total hole character with respect to FEEs. The exact values are listed in Table~\ref{tab:IP_results} for pCCD-based methods and Table S3 in SI for DFT.

The FEEs increase with the degree of doping in both methods (Figure~\ref{fig:HL_FEE}(c–d)). Doping appears to enhance the electronic coupling between the bridge and the acceptor; however, it simultaneously weakens the donating strength of the donor, as reflected by the significantly smaller D$\rightarrow$B* contributions compared to B$\rightarrow$A*. 
A similar trend is observed for the HOMO–LUMO gaps calculated using TD-CAM-B3LYP, where all values are higher than those for the undoped reference system. 
In contrast, the HOMO–LUMO gaps obtained with EOM-pCCD+S do not follow a straightforward trend, with some values lying above and others below the undoped dye.
As shown in Figure~\ref{fig:hole_FEE}, both methods indicate an increase in hole character and first excitation energy with the degree of doping. The trend obtained with CAM-B3LYP is nearly linear and depends primarily on the extent of doping (Figure~\ref{fig:hole_FEE}(b)). In contrast, the EOM-pCCD+S results are more dispersed, showing additional variations that depend on the type of dopant ($\mathrm{{\color{nitrogen}N}}$, $\mathrm{{\color{oxygen}O}}$ or $\mathrm{{\color{sulfur}S}}$). For instance, among the homogeneously tri-doped dyes, the hole character correlates well with the first excitation energy. The $\mathrm{{\color{sulfur}S}{\color{sulfur}S}{\color{sulfur}S}}$ system exhibits the weakest charge transfer and the highest excitation energy, while the $\mathrm{{\color{oxygen}O}{\color{oxygen}O}{\color{oxygen}O}}$ and $\mathrm{{\color{nitrogen}N}{\color{nitrogen}N}{\color{nitrogen}N}}$ systems show a decrease in this energy as the hole character increases, in line with chemical intuition.

Table~\ref{tab:IP_results} collects the IPs, EAs, and charge gaps computed from the pCCD-based methods for all investigated organic dyes. 
We observed that IPs are very similar for all mono-, di-, and tri-doped nitrogen systems. 
However, IP values increase slightly for oxygen- and sulfur-doped systems. 
Specifically, the IPs increase as the doping level changes from mono- to di- to tri-doped systems.
Among them, sulfur-doped organic dyes exhibit the highest IPs compared to their oxygen and nitrogen analogues.
In mixed di-doped dyes where the bridge is doped with one nitrogen and one sulfur atom, placing the nitrogen at positions closer to the acceptor (for example, positions 2 or 3 relative to sulfur at 1 or 2) results in slightly higher IPs (cf. Table~\ref{tab:IP_results}).
In mixed tri-doped dyes, the IPs increase with a higher concentration of sulfur atoms. Specifically, systems doped with one nitrogen atom and two sulfur atoms have higher IPs than those doped with two nitrogen atoms and one sulfur atom.

The electron affinities of nitrogen-, oxygen-, and sulfur-doped systems increase with doping level.
The effect of nitrogen- and mixed-doping has the largest influence on the EAs, with $\mathrm{{\color{nitrogen}N}{\color{nitrogen}N}{\color{nitrogen}N}}$ and $\mathrm{{\color{nitrogen}N}{\color{sulfur}S}{\color{nitrogen}N}}$ have the largest positive EAs. 
The difference between IPs and EAs (known as the HOMO-LUMO gap or charge gap) of the entire DSSC molecule provides a rough approximation to the first excitation energy. 
As we can see from Table~\ref{tab:IP_results} by comparing them to FEE, this is not always the case, with $\mathrm{{\color{nitrogen}N}{\color{nitrogen}N}{\color{nitrogen}N}}$ showing up to a 1 eV difference.  
A slight increase in the first excitation energy (FEE) is observed for both nitrogen and oxygen dopants as the position changes in mono-doped systems (from position 1 to 2 to 3) and in di-doped systems (from (1,2) to (2,3) to (1,3)), as given in Table~\ref{tab:IP_results}.
In contrast, sulfur-doped systems show no such positional dependence, with FEE values remaining consistent across all mono- and di-doped configurations.

\section{Conclusions}
In this work, we used a novel, reliable, and cost-effective computational methodology for tracking directional charge transfer phenomena in $\pi$-conjugated organic compounds.
Our approach leverages the localized nature of pCCD natural orbitals, which enables us to quantify the charge transfer flow in the D-$\pi$-A model compounds consistently.
Specifically, we systematically investigated mono-, di-, and tri-doping strategies with N, O, and S in the bridge of a core molecular structure comprising a carbazole donor and a cyanoacrylic acid acceptor. 
                      
In mono-doped $\pi$-conjugated organic dyes, the total forward charge transfer efficiency from donor-to-bridge-to-acceptor (D$\rightarrow$B$\rightarrow$A) increases progressively as the heteroatom dopant is shifted from the donor-side (position 1) more towards the acceptor-side (position 3) of the bridge.
Nitrogen-doped systems consistently outperform their oxygen- and sulfur-doped counterparts, with the mono-doped nitrogen system at position 3 ($\mathrm{CC{\color{nitrogen}N}}$) identified as the optimal configuration, achieving the highest forward charge transfer efficiency of 31.5\%. 

In pure di-doped systems, a progressive enhancement in  forward charge transfer (D$\rightarrow$B$\rightarrow$A) is observed with nitrogen and oxygen systems when the doping sites change from positions (1,2) to (2,3) to (1,3).
In di-doped nitrogen-sulfur mixed systems, the relative position of heteroatoms critically determines the charge transfer efficiency.
Positioning nitrogen, a strong lone pair donor (at positions 2 or 3) to the right of sulfur (at positions 1 or 2) establishes a more effective electron-transfer pathway. 
The optimal di-doping configuration with heteroatoms at positions 1 and 3 ($\mathrm{{\color{nitrogen}N}C{\color{nitrogen}N}}$) shows the highest total forward charge transfer (39.2\%).

Among tri-doped molecular systems, the most efficient organic dye is a homogeneously doped system with three nitrogen atoms ($\mathrm{{\color{nitrogen}N}\mathrm{{\color{nitrogen}N}\mathrm{\color{nitrogen}N}}}$) achieves the highest forward charge transfer efficiency of 42.6\%, significantly outperforming both the oxygen (34.0\%) and sulfur (25.8\%) counterparts.
In tri-doped mixed configurations, system doped with two sulfur atoms and one nitrogen atom shows maximum charge transfer (35.8\%) when sulfur occupies positions 1 and 2 and nitrogen is at position 3 ($\mathrm{{\color{sulfur}S}{\color{sulfur}S}{\color{nitrogen}N}}$). However, an organic dye containing two doped nitrogen atoms and one sulfur atom exhibit the largest charge transfer (40.3\%) when nitrogens are placed at positions 1 and 3 and sulfur is at position 2 ($\mathrm{{\color{nitrogen}N}{\color{sulfur}S}{\color{nitrogen}N}}$).

These results confirm that systematically increasing the concentration of nitrogen doping from mono- to di- and then to tri-doping enhances the charge transfer, with the fully tri-doped nitrogen bridge representing the optimal configuration for maximizing charge transfer in organic dye sensitizers.

Our results suggest a different mechanism for the charge-transfer process. The excitation occurs primarily on the bridge, and the charge is transferred subsequently to the acceptor through the conjugated $\pi$-system. The remaining hole is then recombined via charge transfer from the donor, consistent with the higher probability of D$\rightarrow$B* compared to B*$\leftarrow$A charge transfer, leaving a hole on the donor domain. However, to fully exploit the charge-donating potential of the donor and achieve efficient charge separation, considering an isolated dye is not enough, and a semiconductor needs to be introduced to model the charge transfer process in dye-sensitized solar cells.

\section*{Conflicts of interest}
There are no conflicts to declare.

\begin{suppinfo}
The following data are available free of charge.
\begin{itemize}
  \item Dihedral angles between the donor (D) and the bridge (B) and between the bridge (B) and acceptor (A) for 33 investigated dyes. 
  \item Calculated HOMO, LUMO, HOMO-LUMO gap, and first excitation energies using CAM-B3LYP/cc-pVDZ.
  \item Contribution of holes and particles for each domain with CAM-B3LYP/cc-pVDZ using TheoDORE.

\end{itemize}

\end{suppinfo}

\section*{Data Availability Statements}
The data underlying this study are available in the published article and its Supporting Information.
The released version of the PyBEST code is available on Zenodo at \url{https://zenodo.org/records/10069179} and on PyPI at \url{https://pypi.org/project/pybest/}.

\begin{acknowledgement}
M.G.~acknowledges financial support from the SONATA research grant from the National Science Centre, Poland (Grant No. 2023/51/D/ST4/02796). 
P.T. acknowledges financial support from the PRELUDIUM BIS research grant from the National Science Centre, Poland (grant no. 2023/50/O/ST4/00353). 
The research leading to these results has received funding from the Norway Grants 2014--2021 via the National Centre for Research and Development.
We acknowledge that the results of this research have been achieved using the DECI resource Bem (Grant No.~412 and Grant No.~411) based in Poland at Wroclaw Centre for Networking and Supercomputing (WCSS, http://wcss.pl) with support from the PRACE aisbl.
This work was completed in part at the Poland Open Hackathon, part of the Open Hackathons program. 
The authors acknowledge OpenACC-Standard.org for their support.
We gratefully acknowledge Polish high-performance computing infrastructure PLGrid (HPC Center: ACK Cyfronet AGH) for providing computer facilities and support within computational grant no.~PLG/2024/017774.
Funded/Co-funded by the European Union (ERC, DRESSED-pCCD, 101077420 ).
Views and opinions expressed are, however, those of the author(s) only and do not necessarily reflect those of the European Union or the European Research Council. Neither the European Union nor the granting authority can be held responsible for them. 
 
\end{acknowledgement}

\bibliography{acs}
\end{document}